\documentclass[preprint,12pt]{elsarticle}

\usepackage{amssymb}
\usepackage{color}

\journal{NIM A}

\begin{document}

\begin{frontmatter}



\title{Investigating the image lag of a scientific CMOS sensor in X-ray detection}

\author[inst1,inst2]{Qinyu Wu}
\author[inst1,inst2]{Zhixing Ling\corref{cor1}}
\ead{lingzhixing@nao.cas.cn}
\author[inst1,inst2]{Chen Zhang}
\author[inst3]{Quan Zhou}
\author[inst3]{Xinyang Wang}
\author[inst1,inst2]{Weimin Yuan}
\author[inst1,inst2,inst4]{Shuang-Nan Zhang}

\affiliation[inst1]{organization={National Astronomical Observatories, Chinese Academy of Sciences},
            addressline={20A Datun Road, Chaoyang District}, 
            city={Beijing},
            postcode={100101}, 
            country={China}}
\affiliation[inst2]{organization={School of Astronomy and Space Science, University of Chinese Academy of Sciences},
            addressline={19A Yuquan Road, Shĳingshan District}, 
            city={Beijing},
            postcode={100049}, 
            country={China}}
\affiliation[inst3]{organization={Gpixel Inc.},
            addressline={No. 588 Yingkou Road, Economical and Technological Development Zone}, 
            city={Changchun},
            postcode={130033}, 
            country={China}}
\affiliation[inst4]{organization={Institute of High Energy Physics, Chinese Academy of Sciences},
            addressline={19B Yuquan Road, Shĳingshan District}, 
            city={Beijing},
            postcode={100049}, 
            country={China}}
\cortext[cor1]{Corresponding author}

\begin{abstract}
In recent years, scientific CMOS (sCMOS) sensors have been vigorously developed and have outperformed CCDs in several aspects: higher readout frame rate, higher radiation tolerance, and higher working temperature. For silicon image sensors, image lag will occur when the charges of an event are not fully transferred inside pixels. It can degrade the image quality for optical imaging, and deteriorate the energy resolution for X-ray spectroscopy. In this work, the image lag of a sCMOS sensor is studied. 
To measure the image lag under low-light illumination, we constructed a new method to extract the image lag from X-ray photons. The image lag of a customized X-ray sCMOS sensor GSENSE\-1516\-BSI is measured, and its influence on X-ray performance is evaluated. The result shows that the image lag of this sensor exists only in the immediately subsequent frame and is always less than $0.05 \%$ for different incident photon energies and under different experimental conditions. The residual charge is smaller than $0.5\ \rm{e^-}$ with the highest incident photon charge around $8\ \rm{ke^-}$. Compared to the readout noise level around $3\ \rm{e^-}$, the image lag of this sensor is too small to have a significant impact on the imaging quality and the energy resolution. The image lag shows a positive correlation with the incident photon energy and a negative correlation with the temperature. However, it has no dependence on the gain setting and the integration time. These relations can be explained qualitatively by the non-ideal potential structure inside the pixels. This method can also be applied to the study of image lag for other kinds of imaging sensors.
\end{abstract}




\end{frontmatter}


\section{Introduction}
\label{sec:intro}  
Silicon image sensors, including charge-coupled devices (CCDs) and Complementary Metal Oxide Semiconductor (CMOS) sensors, have been widely used for optical and X-ray applications in the fields of medical imaging, industrial radiography, nuclear physics and astronomical research. For examples, since the 1990s, CCDs have been used as soft X-ray imaging and spectroscopy detectors in many synchrotron light sources \citep{strauss1988ccd,clarke1994ccd,doering2011ccd}, and in space X-ray astronomical missions, such as ASCA \citep{burke1994ccd}, Chandra \citep{garmire2003advanced} and XMM-Newton \citep{struder2001european}. In recent years, scientific CMOS (sCMOS) sensors have progressed rapidly and have outperformed CCDs in several aspects: high readout frame rate, high radiation tolerance, and relaxed working temperature requirement, making the sCMOS sensor a favorable choice in both ground experiments and space missions. In August 2022, the performance of a customized 6 cm $\times$ 6 cm large-format X-ray sCMOS sensor has been successfully verified in orbit by the Lobster Eye Imager for Astronomy (LEIA) onboard the SATech-01 satellite \citep{zhang2022first, ling2023thelobster}. In the future, missions such as Einstein Probe (EP) \citep{yuan2018einstein,yuan2022the} and THESEUS \citep{heymes2020development} will use sCMOS sensors as focal plane detectors.

As a new type of imaging sensors, the performance of sCMOS sensors has been studied less extensively than that of mature CCDs. Since 2015, We have tested several on-shelf sCMOS sensors and evaluated their X-ray performance and capabilities of X-ray imaging and spectroscopy \citep{wang2019characterization, wang2022design_Al, hsiao2022xray}. Crosstalk was also found in a standard GSENSE\-400\-BSI sensor \citep{ling2021correlogram}. In this work, we will focus on the image lag of the sCMOS sensor. CCD and CMOS sensors can detect the input signal (including optical, UV, and X-ray photons and charged particles) by converting it to photon-electrons, and collect them into the nearby pixels. If the photon-generated electrons are not completely transferred and collected in one frame, image lag will occur as the residual electrons are transferred in the subsequent frames. Image lag is important for silicon image sensors and is closely related to the pixel's physical structure. For optical applications, severe image lag will leave a ghost image after an exposure to a bright target. For X-ray applications, image lag can deteriorate the energy resolution. Many works have been done to reduce image lag by optimizing the pixel design \citep{teranishi1984an, yu2010two, xu2013image, miyauchi2014pixel,lofthouse2018image, stefanov2022a}. The image lag of a silicon image sensor can be measured with an optical flash of a LED \citep{xu2013image, miyauchi2014pixel, lofthouse2018image}, 
and the lag is typically less than 1$\%$ for modern CCD and CMOS sensors. However, for some specialized sensors, such as aluminum-coated sCMOS sensors designed for X-ray applications \citep{wang2022design_Al}, the optical method cannot be applied and new methods have to be adopted. X-ray photons can produce single-pixel events on a sCMOS sensor and make it possible to assess the image lag performance on the pixel level.

Cooperating with Gpixel Inc., our lab has customized a 6 cm $\times$ 6 cm large-format sCMOS sensor, which is available in 2019. This sCMOS sensor, named GSENSE\-1516\-BSI \citep{WU2022}, is optimized for soft X-ray detection. It has an array of $4096\times4096$ pixels with a pixel size of $\rm{15\ \mu m}$. The epitaxial layer is $\rm{10\ \mu m}$ thick, and it is fully depleted. The dark current of this sensor is lower than 0.02 $\rm{e^-}$/pixel/s and the readout noise is lower than 5 $\rm{e^-}$ at $\rm{-30 ^{\circ}\!C}$. The nominal energy resolution is about 180 eV at 5.9 keV. We have demonstrated \citep{wu2023improving} that, by making pixel-level gain correction, its energy resolution can be improved to 124.6 eV at 4.5 keV and 140.7 eV at 6.4 keV at room temperature. A level of extra noise of $\sim 7\ \rm{e^-}$ is needed to explain the difference from the theoretical limit, and image lag is a candidate source of the noise. To investigate the image lag of this X-ray sCMOS sensor and assess its effect on the energy resolution, we developed a new method using X-ray measurement and carried out a series of experiments.

The experimental setup is introduced in Section~\ref{sec:exp_setup}; the measurement method is described in Section~\ref{sec:data_reduce}; the results about the image lag are shown and discussed in Section~\ref{sec:results_and_discuss}; and the conclusions are summarized in Section~\ref{sec:conclu}.

\section{Experimental Setup}
\label{sec:exp_setup}
\begin{figure}
\begin{center}
\begin{tabular}{c}
\includegraphics[width=0.7\textwidth]{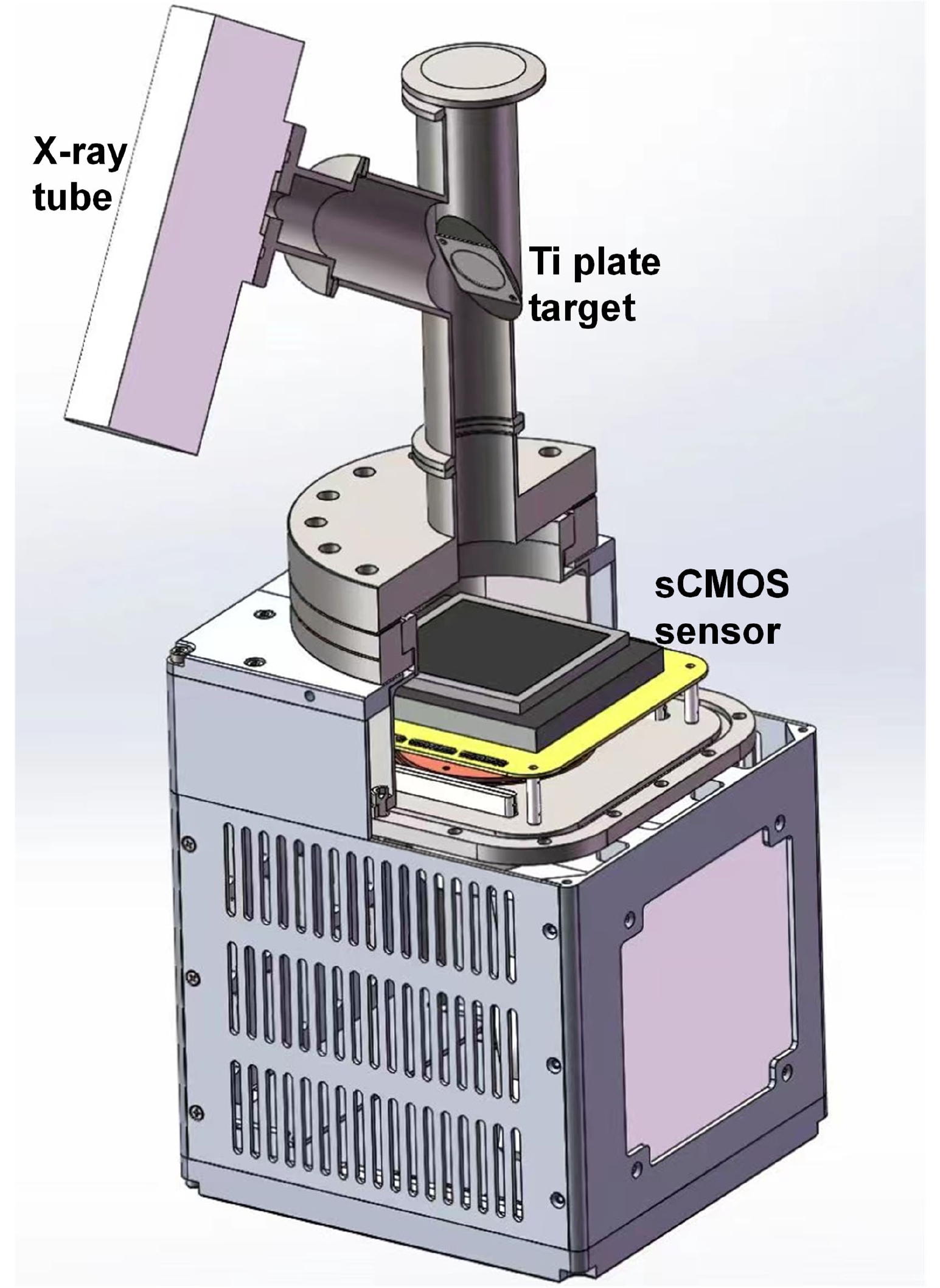}
\end{tabular}
\end{center}
\caption 
{ \label{fig:exp_device}
The experimental setup. The center Ti plate target in this setup, illuminated by the X-ray tube at left side, can reduce the initial X-ray flux and produce secondary X-ray photons. These photons are then recorded by the sCMOS sensor in the camera at the bottom.} 
\end{figure}

The experimental device for image lag measurement is shown in Fig.~\ref{fig:exp_device}. An Oxford Jupiter 5000 X-ray tube with a Ti anode is placed at the top left side of the setup. At the center, a Ti plate target is installed to produce secondary X-rays and to reduce the initial X-ray flux. When the X-ray tube is powered on, an X-ray continuum spectrum with emission lines from the plate is produced. These X-rays are then recorded by a camera at the bottom. The camera is made of a GSENSE\-1516\-BSI sCMOS sensor, readout electronics, and temperature control structures \citep{wang2022design}. The camera has two working modes: the image mode, in which raw images are read out and saved into a disk directly; and the event mode, in which X-ray events are extracted from raw images in real time and only such events are saved into the disk. The event mode can greatly reduce the bandwidth needed for high-speed exposures, especially at the highest frame rate of 20 Hz. 

\begin{figure}
\begin{center}
\begin{tabular}{c}
\includegraphics[width=0.8\textwidth]{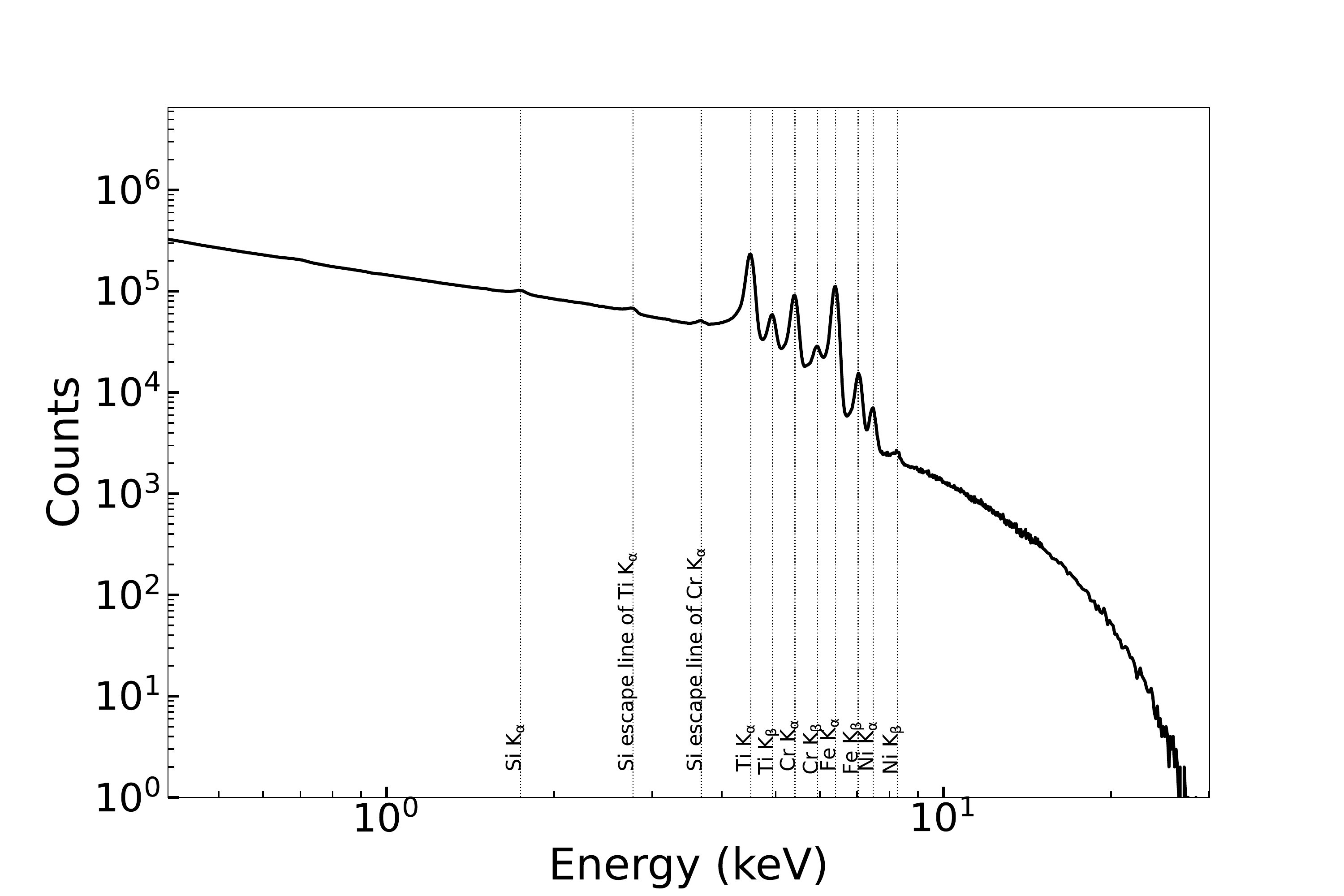}
\end{tabular}
\end{center}
\caption 
{ \label{fig:spectrum}
A typical spectrum obtained by the sCMOS sensor. X-rays close to 30 keV can be seen. Characteristic lines of not only Ti, but also Cr, Fe and Ni can be found in the spectrum, which come from the stainless steel of the surrounding structures of the device.} 
\end{figure}

In this study, the X-ray tube is operated under a current of 0.12 mA and a high-voltage of 30kV. Therefore, X-rays close to 30 keV can be seen in the final spectrum, as shown in Fig.~\ref{fig:spectrum}. However, at this high energy, the quantum efficiency of the sensor is pretty low \citep{desjardins2020backside,menk2022on}, and only a small number of the incident photons can be captured by the sensor. Characteristic lines of not only Ti, but also Cr, Fe and Ni can be found in the spectrum, which come from the stainless steel of the surrounding structures of the device. The countrate is controlled to be about 20000 counts per frame to make the pixel occupancy of each frame less than 0.2\%. In a vacuum of around $10^{-3}$ Pa, a series of experiments were carried out with different gain settings and different integration times, and at different temperatures, ranging from $\rm{-30 ^{\circ}\!C}$ to $\rm{20 ^{\circ}\!C}$

Under each of the experiment conditions, dark exposures without X-ray illumination are first taken to obtain the raw background images. The experiments are then conducted with the X-ray tube powered on. For the following data reduction, the continuous raw images captured by the sCMOS sensor have to be stored, so we make the camera run in the image mode.

\section{Measurement Method}
\label{sec:data_reduce}
\begin{figure}
\begin{center}
\begin{tabular}{c}
\includegraphics[width=1.0\textwidth]{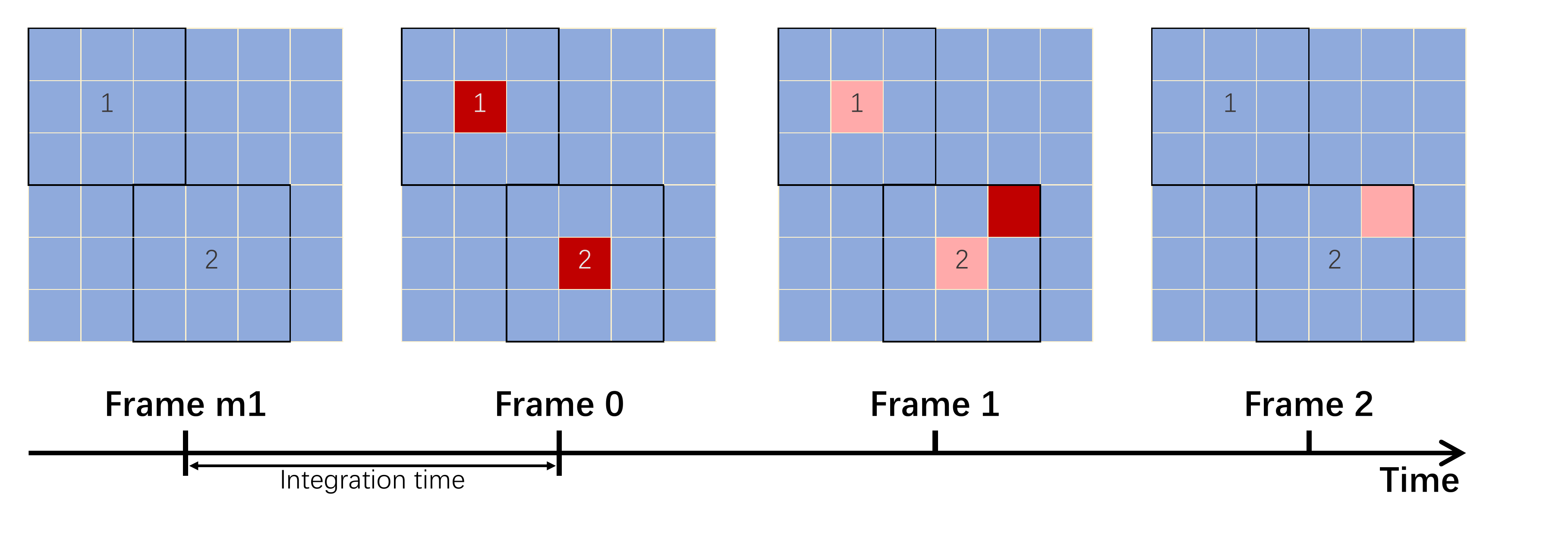}
\end{tabular}
\end{center}
\caption 
{ \label{fig:frame_scheme}
Schematic diagram of the frames. Frame m1, 0, 1 and 2 are four frames in a continuous exposure. The time between two frames is the integration time. As examples, two over-threshold pixels are given in frame 0. For event 1 in the upper-left part of each panel, there is no other incident photon in frame m1, frame 1 and frame 2. Therefore, this event will pass the selection process. But for event 2 in the bottom part of each panel, another incident photon is detected on an adjacent pixel in frame 1, and therefore event 2 cannot be used for the image lag calculation.} 
\end{figure}

Under each experimental condition, an averaged background image is calculated from dark exposures. Then the X-ray exposure data is processed frame by frame. For each frame (frame 0), we examine the preceding frame (frame m1) and the two subsequent frames (frame 1 and frame 2), as shown in Fig.~\ref{fig:frame_scheme}. After subtracting the background image from the X-ray image, we extract all pixels above a threshold of around 335 eV or 92 $\rm{e^{-}}$ in frame 0 to eliminate false events caused by noise and keep those real events from incident X-ray photons. 

To obtain an unbiased image lag, we record the corresponding pixel values in frame m1, frame 1 and frame 2. However, these pixel values may be affected by another incident X-ray photon and cannot represent the real residual signal. Therefore, an additional selection process is carried out on each over-threshold pixel of frame 0: In frame m1, frame 1 and frame 2, if the values of the pixel and its eight adjacent pixels are all below the threshold around 335 eV, which means that this pixel is not affected by photons in these three frames, then the value difference of frame 1 and frame m1 is a reliable measurement of the residual signal or the image lag in frame 1, and similarly the difference between frame 2 and frame m1 is a reliable measurement for frame 2. Finally, around 6\% of these over-threshold pixels are discarded, corresponding well to the occupancy around 0.2\%. Fig.~\ref{fig:frame_scheme} gives two examples of this selection process. For event 1 in the upper-left part of each panel, there is no other incident photon in frame m1, frame 1 and frame 2. Therefore, this event will pass the selection process. But for event 2 in the bottom part of each panel, another incident photon is detected on an adjacent pixel in frame 1, and therefore event 2 cannot be used for the image lag calculation. 

\begin{figure}
\begin{center}
\begin{tabular}{c}
\includegraphics[width=0.8\textwidth]{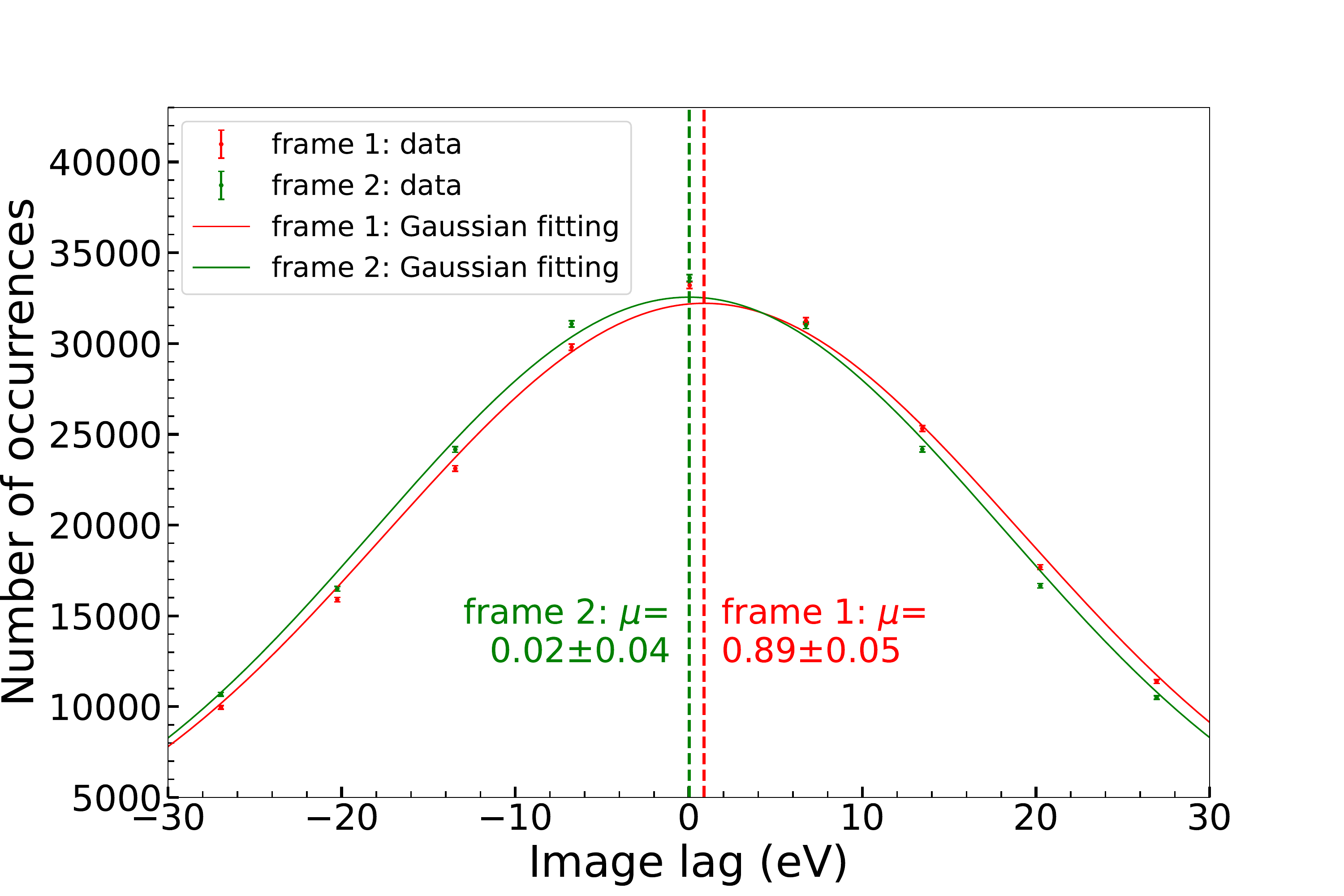}
\end{tabular}
\end{center}
\caption 
{ \label{fig:fit_example}
The distribution of the measurement results of the image lag for frame 1 and frame 2, respectively. The Gaussian fitting of the distributions are also shown. Errors are given with a confidence level of 1 sigma. This figure is for incident photons around 12 keV. The temperature is $\rm{-30 ^{\circ}\!C}$ and the gain register is 7.5. For frame 2, the second subsequent frame of frame 0, the mean value $\mu$ is around 0, indicating that there is no image lag in frame 2 under this experimental condition. For frame 1, the immediately subsequent frame of frame 0, $\mu$ gives a image lag of around 0.89 eV, or 0.24 $\rm{e^{-}}$. The direct average of the histogram gives a value of 0.26 $\rm{e^{-}}$, which is close to the Gaussian fitting result.}
\end{figure}

For each selected pixel on each frame, we can get a single measurement value of the image lag. We perform this selection and measurement for all over-threshold pixels on all frames, so a large number of measurement results of the image lag are obtained. The overall distribution of these image lags for frame 1 and frame 2, respectively, are shown in Fig.~\ref{fig:fit_example}. A Gaussian function is used to fit the histograms, and the mean value $\mu$ of the Gaussian fit gives the image lag level of the sensor. The direct average of the histogram gives a similar result.

\section{Results and Discussion}
\label{sec:results_and_discuss}
Our results show that, under all experimental conditions adopted, the image lag of frame 2, the second subsequent frame of a given frame, is statistically consistent with zero. Fig.~\ref{fig:fit_example} gives such an example under a specific experimental condition. We conclude that there is no measurable image lag in the second subsequent frame. 

\begin{figure}
\begin{center}
\begin{tabular}{c}
\includegraphics[width=0.8\textwidth]{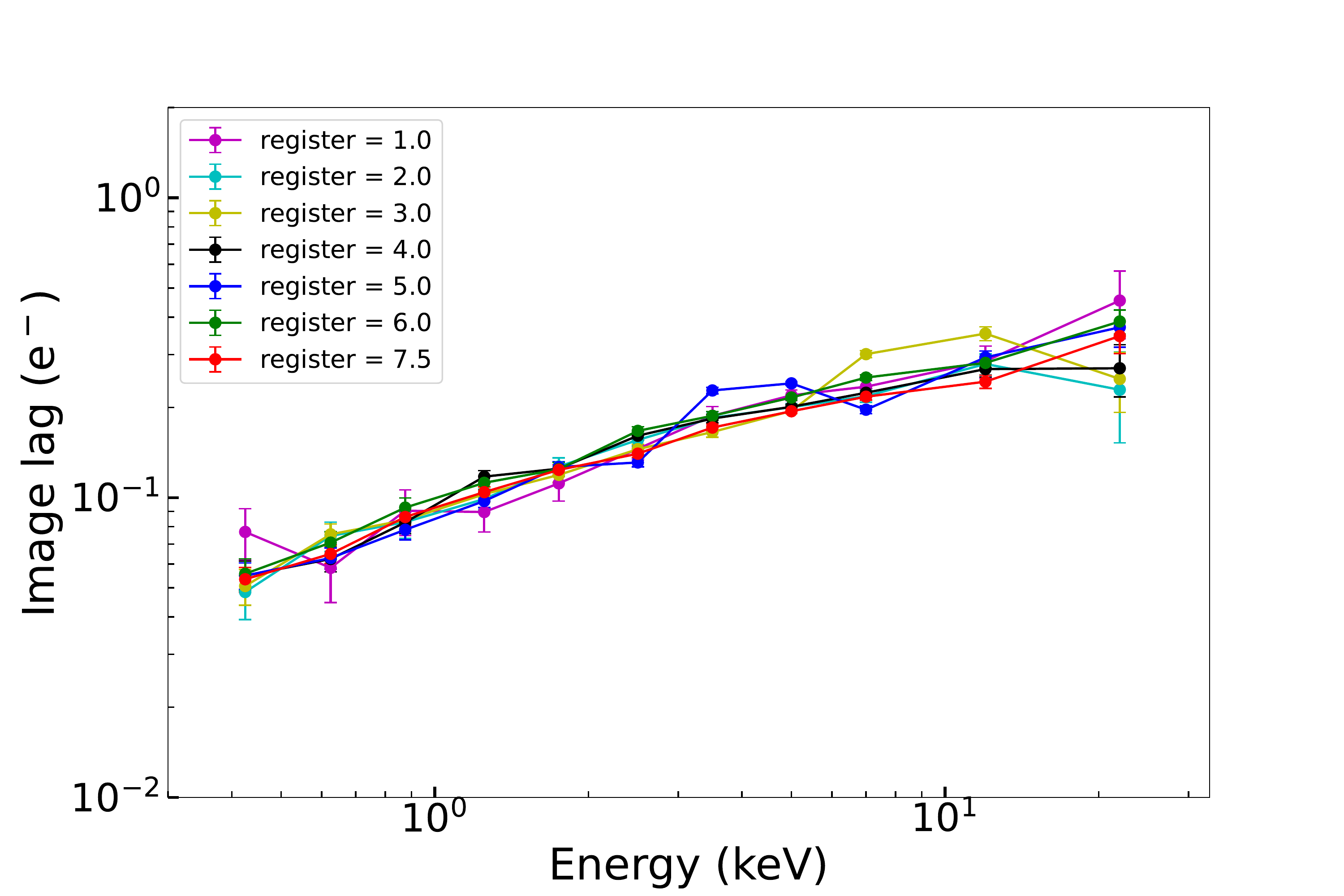}
\end{tabular}
\end{center}
\caption 
{ \label{fig:lag_reggain}
The image lag versus the incident photon energy, under different gain settings. The conversion gain of this sCMOS sensor can be tuned from about 6.5 eV/DN to around 45 eV/DN, corresponding to a gain register value of 7.5 to 1. The image lag increases as the incident energy increase, and it shows no dependence on the gain setting. The integration time is 50 ms and the temperature is kept at $\rm{-30 ^{\circ}\!C}$.} 
\end{figure}

However, for the immediately subsequent frame (frame 1), the image lag is present clearly. We divide the data into several groups by the energy of the incident photon to study the dependence of the image lag on the energy of incident photons. Fig.~\ref{fig:lag_reggain} shows the image lags measured versus the energies of incident photons. A positive correlation is clearly present,indicating that the image lag is positively correlated to the number of electrons in the original frame. However, as the photon energy increases, the percentage of the residual charge among the total deposited charge decreases.

The effect of the gain setting on the image lag is also examined. The conversion gain of this CMOS sensor can be tuned from about 6.5 eV/DN to around 45 eV/DN, corresponding to a gain register value of 7.5 to 1. Fig.~\ref{fig:lag_reggain} also shows the energy-dependent image lags under different gain register values. There is no obvious dependence of the image lag on the gain setting, indicating that the image lag of this sensor does not come from the readout circuits and the digital electronics.

\begin{figure}
\begin{center}
\begin{tabular}{c}
\includegraphics[width=0.8\textwidth]{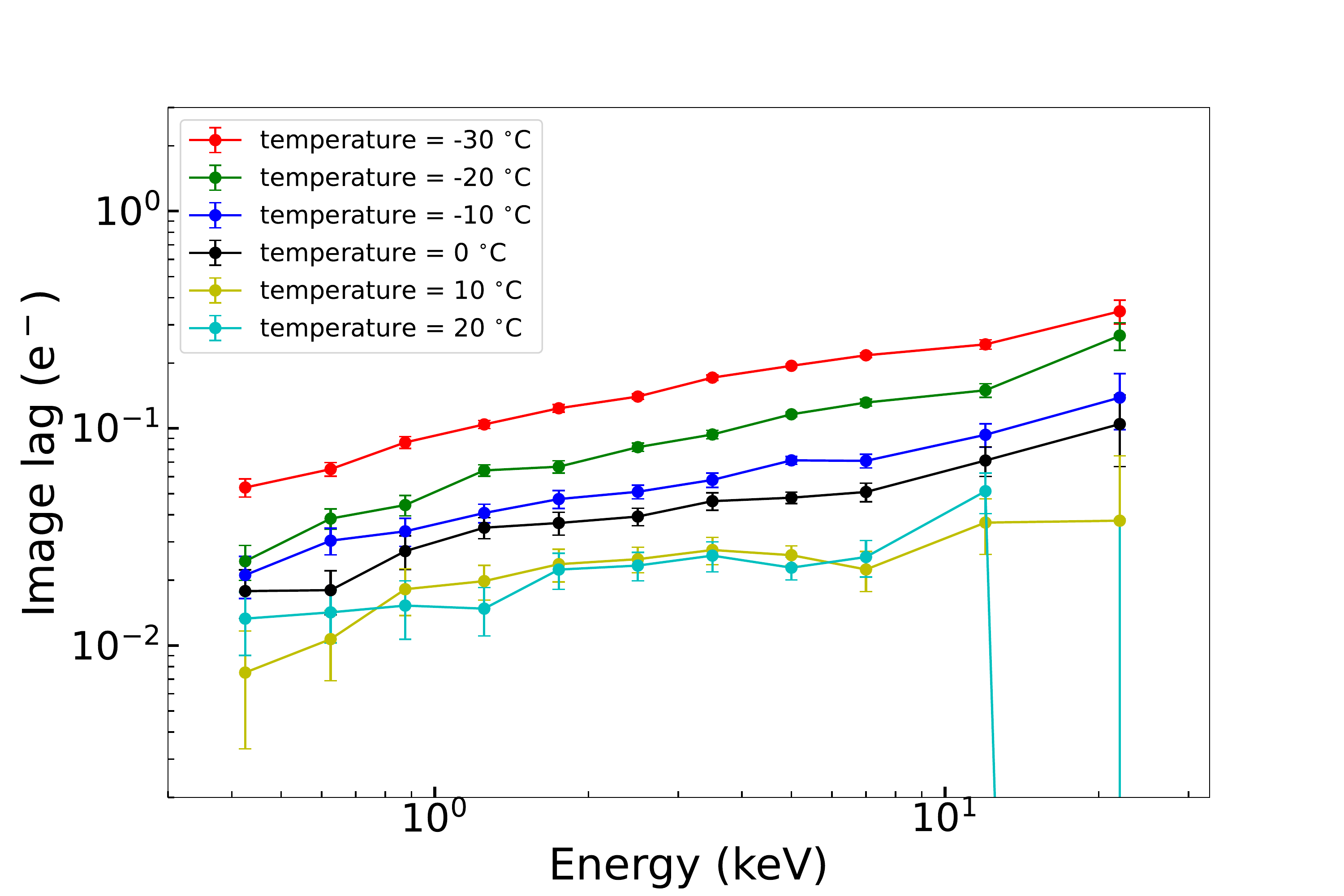}
\end{tabular}
\end{center}
\caption 
{ \label{fig:lag_temp}
The image lag versus the incident photon energy, under different temperatures. The image lag decreases as the temperature increases. Above $\rm{10 ^{\circ}\!C}$, the image lag becomes small enough and the statistical uncertainty starts to dominate the measurement; therefore, the variation among the data points becomes obvious in a log scale plot. The gain register is set to 7.5 and the integration time is 50 ms.} 
\end{figure}

The image lags are also measured at different temperatures. As shown in Fig.~\ref{fig:lag_temp}, the image lag decreases significantly as the temperature increases. The positive correlation trend between the lag and the energy is kept the same at these temperatures. Above $\rm{10 ^{\circ}\!C}$, the image lag becomes small enough and the statistical uncertainty starts to dominate the measurement; therefore, the variation among the data points becomes obvious in a log scale plot.

\begin{figure}
\begin{center}
\begin{tabular}{c}
\includegraphics[width=0.8\textwidth]{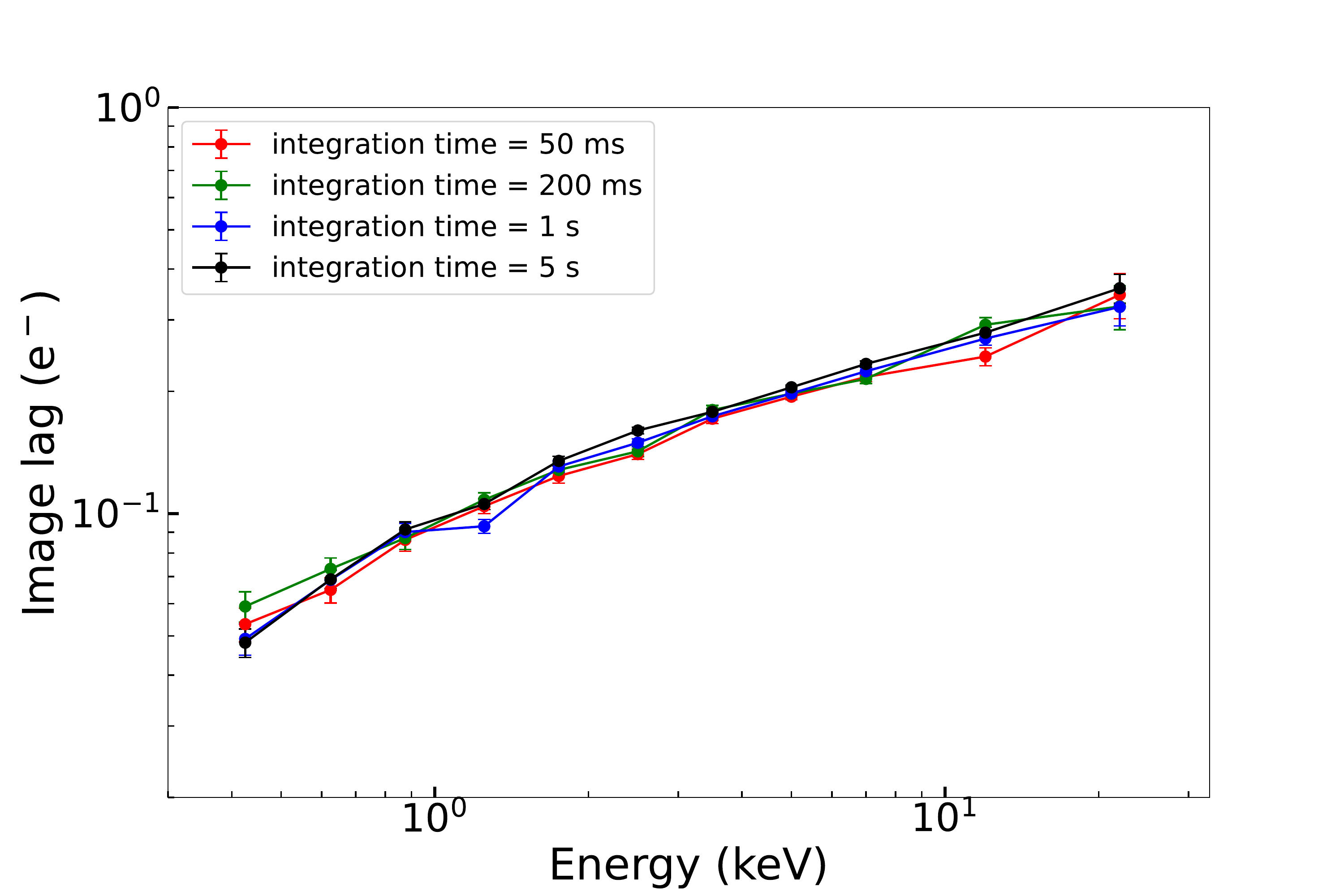}
\end{tabular}
\end{center}
\caption 
{ \label{fig:lag_inttime}
The image lag versus the incident photon energy, under different integration times. The image lag shows no dependence on the integration time. The gain register is set to 7.5 and the temperature is kept at $\rm{-30 ^{\circ}\!C}$.} 
\end{figure}

The image lags under different integration times, ranging from 50 ms to 5 s, are shown in Fig.~\ref{fig:lag_inttime}. Despite a change of the integration time by a factor of ten, the image lag remains the same. This shows that the integration time of the sensor has no measurable effect on the image lag and proves that the image lag is not related to the charge collection process.

\begin{figure}
\begin{center}
\begin{tabular}{c}
\includegraphics[width=0.95\textwidth]{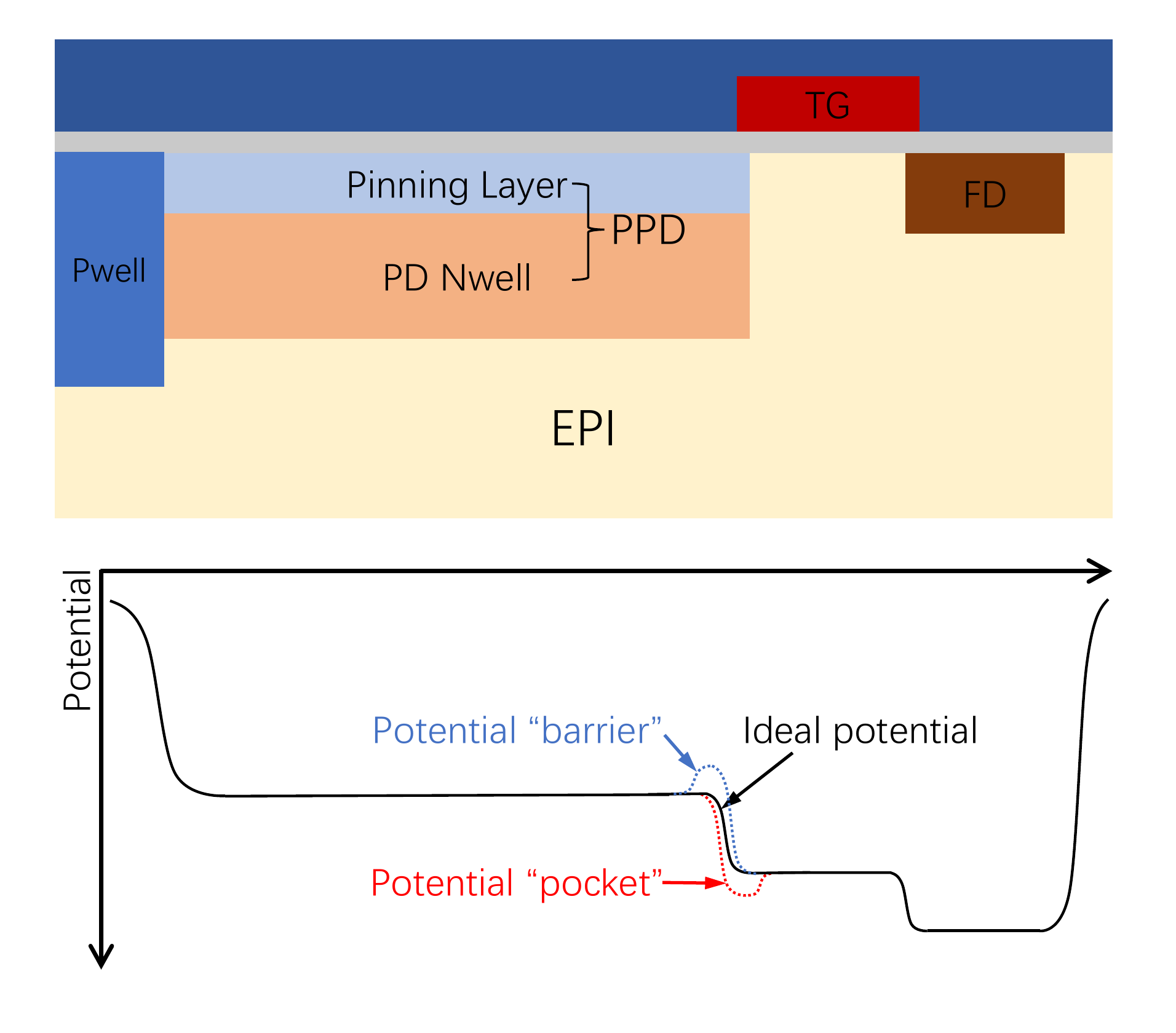}
\end{tabular}
\end{center}
\caption 
{ \label{fig:pixel_scheme}
A typical pixel design of a 4T CMOS sensor (top) and its possible electric potential distributions during the charge transfer process. The PPD is the place where electrons are collected. After the collection process, the TG will be turned on and electrons will flow down to the FD. Then the charge in FD will be converted to voltage signals and read out. If the potential has a ``barrier" or a ``pocket", some electrons may be blocked or trapped, and image lag will present.} 
\end{figure}

To summarize, the image lag is correlated positively with incident photon energy and inversely with the temperature, but is independent of the gain setting and integration time of the experiment. There have been extensive discussions on the causes of image lag. They are usually thought to be related to the pixel design of a sensor \citep{teranishi1984an, yu2010two, xu2013image, lofthouse2018image}. Fig.~\ref{fig:pixel_scheme} shows the typical pixel design and its possible electric potential distribution of a 4-Transistor CMOS sensor. The pinned photodiode (PPD) is the place where electrons are collected. After the collection process, the transfer gate (TG) will be turned on and the electrons will flow down to the floating diffusion (FD) node. Then the charge in FD will be converted to voltage signals and read out. If the potential has a ``barrier" or a ``pocket" at the edge of the PPD and the opened TG, as shown in Fig.~\ref{fig:pixel_scheme}, electrons can be blocked by the barrier or trapped by the pocket, and cannot reach the FD during one charge transfer process. In this case, the image lag is present. As the temperature increases, the thermal energies of the electrons increase and fewer electrons will be trapped by the small well. Therefore, the image lag is smaller at higher temperatures. The gain settings and the charge integration process have nothing to do with the charge transfer process, and therefore, have no effect on the image lag. As the photon energy increases, more electrons will be produced, and it is reasonable that more electrons will be blocked or trapped. However, as the photon energy increases, the percentage of the trapped charge among the total deposited charge decreases because the effective barrier or pocket is weakened by the already trapped charges. These are consistent with our experimental results.

The image lags of GSENSE\-1516\-BSI are always less than $0.05 \%$ for different incident photon energies and under experimental conditions. The residual charge is smaller than $0.5\ \rm{e^-}$ with the highest incident energy around 30 keV or $8\ \rm{ke^-}$. This value is much smaller than the readout noise around $3\ \rm{e^-}$\citep{wu2023improving}, indicating that the image lag of this sensor is negligible and has no noticeable effect on the energy resolution.

In this work, the image lag of this sCMOS sensor is measured with X-rays. We remark that this method can be applied to image lag measurements using other means of radiation such as electrons and protons. We measured the image lag of the GSENSE\-1516\-BSI sensor with cosmic rays in a normal environment, and the results are similar to the results of X-rays, demonstrating the applicability of our method.

\section{Conclusions}
\label{sec:conclu}
Scientific CMOS sensors have been widely used in ground X-ray experiments and space X-ray missions due to their excellent performance. As an important factor, the image lag resulting from incomplete charge transfer in the pixels can deteriorate the image quality and the energy resolution of sCMOS sensors. In this work, we propose a new method to extract the image lag from X-ray photons. This method is applied to an extensive study of the image lag for a customized X-ray sCMOS sensor GSENSE\-1516\-BSI, for a range of incident energies and under various experimental conditions. It is found that the image lag of this sensor is always less than $0.05 \%$, and the residual charge is smaller than $0.5\ \rm{e^-}$ with the highest incident energy around $8\ \rm{ke^-}$. This low value indicates that the image lag of this sensor has negligible effects on the imaging and energy resolution. The image lag increases slightly with the increase of the incident energy and with the decrease of the temperature. However, it shows no correlation with the gain setting and the integration time of the sensor. These can be explained qualitatively by the non-ideal potential scheme inside the pixel, lending credence to the pixel-level origin of the image lag. In principle, this method can be applied to not only X-ray radiation but other charged particles, showing the generality of our method.
\\ \hspace*{\fill} \\
\noindent \textbf{Funding:} This work is supported by the National Natural Science Foundation of China (grant No. 12173055) and the Chinese Academy of Sciences (grant Nos. XDA15310000, XDA15310100, XDA15310300, XDA15052100).

%

 \bibliographystyle{elsarticle-num} 
 \bibliography{report.bib}





\end{document}